\documentclass[3p,times]{elsarticle}

\usepackage{ecrc}
\usepackage{amsmath}
\usepackage{graphics}
\newcommand{\cwr}[1]{c^{Wr}_{#1}}

\volume{00}

\firstpage{1}

\journalname{Physics Procedia}

\runauth{K. Kampf}

\jid{phpro}

\jnltitlelogo{Physics Procedia}

\usepackage{amssymb}
\usepackage[figuresright]{rotating}

\begin{document}

\begin{frontmatter}



\title{On decays of light unflavoured pseudoscalar mesons}
\author{Karol Kampf}
\ead{karol.kampf@thep.lu.se}
\address{Department of Astronomy and Theoretical Physics, Lund
University,\\S\"olvegatan 14A, SE 223-62 Lund, Sweden.}
\address{Charles University, Faculty of Mathematics and Physics,\\
V Hole\v{s}ovi\v{c}k\'ach 2, Prague, Czech Republic}

\dochead{\small\begin{flushright}
LU TP 11-35
\end{flushright}\large Physics of Fundamental Symmetries and Interactions –--
PSI2010}

%\title{}


%\author{}

%\address{}

\begin{abstract}
The ongoing and planned experimental activities with direct reference to light
unflavoured pseudoscalar mesons motivate a new theoretical
study regarding their properties.
An overview including details on new precise calculations is presented.

\end{abstract}

\begin{keyword}
Chiral symmetries \sep
Decays of $\pi$ mesons \sep
Light mesons
\end{keyword}

\end{frontmatter}

% \linenumbers

\section{Introduction}
%\label{}
The subjects of this work, light pseudoscalar mesons, play a prominent role in
hadronic physics. Our even more focused interest into the unflavoured
particles, namely $\pi^0$, $\eta$ and eventually $\eta'$, is motivated by the
wish to avoid a discussion on $K^0$ decays. Of course, not because they are not
interesting, but such decays violate hypercharge conservation and are suppressed
by Fermi coupling constant $G_F^2$, meaning that studying only unflavoured ones
enables to reduce standard model to QCD and simplifies the problem. For
studying QCD at low energy region, in our case enlarged at most only by QED
corrections, a standard tool successfully developed in recent years is called
chiral perturbation theory (ChPT). In this contribution we will mainly discuss
properties of $\pi^0$ as many radiative $\eta$ decays are technically very
similar and what one obtains for $\pi^0$ can be simply converted also for $\eta$
decay prediction. On the other hand, $\pi^0$ being the lightest meson cannot
decay into other hadronic states, therefore a hadronic discussion for $\eta$
decays
is inevitable. As an important example of such processes we will briefly mention
$\eta\to3 \pi$ decays. 

The decay modes of $\pi^0$ were subjects of many experiments in the past
(including e.g.
SINDRUM coll. at PSI), present (e.g. KTeV, or PrimEx at JLab) or future (NA62 at
CERN). Using the conserved vector current hypothesis we can connect the vector
form
factor (i.e. charged pion) to the lifetime of the neutral pion (cf. PIBETA
 \cite{bychkov}). Experiments have reached (or plan to reach) a level
of precision which
makes it mandatory to reopen previous theoretical calculations to achieve
appropriate order (NLO or NNLO). This can, on one hand, help us to verify and
fix the underlying structure of the low energy effective theory of QCD - ChPT
(e.g.
pion decay constant, low energy constants, power-counting, etc.). On the other
hand it can set a framework for studying new physics beyond the SM (e.g. KTeV's
discrepancy with a theory for $\pi^0 \to e^+e^-$). 

As mentioned the $\eta$ meson can be treated
technically very similarly. However, due to its mass one can also study
different leptonic variants and combination (for example $\mu$ in place of $e$)
and last but not least its
hadronic decays. This can provide us with important information on isospin
breaking effects and again test the internal consistency of ChPT.

We will mainly focus on four most important
allowed decay modes of $\pi^0$: $\gamma\gamma$, $e^+e^-\gamma$, $e^+e^-e^+e^-$,
$e^+e^-$ (with branching ratios \cite{pdg}: $0.98798(32)$, $0.01198(32)$,
$3.14(30)\times 10^{-5}$, $6.46(33)\times 10^{-8}$, respectively). For this
purpose one can use two-flavour chiral perturbation theory (ChPT, for a review
see e.g.\cite{bijnenshere}) which can simply incorporate corrections to the
current
algebra result attributed either to $m_{u,d}$ masses or electromagnetic
corrections with other effects hidden in the low energy constants (LECs),
denoted by $c_i^W$ at next-to-leading order (NLO).
However, phenomenologically richer $SU(3)$ ChPT must be also employed in order
to obtain a numerical prediction. This is especially true for the studied
anomalous processes as in this case the initial symmetry for the two flavour
case must be extended and the number of monomials in $SU(2)$ increases
\cite{bijnens01}.

We will not limit our focus only on ``standard on-shell''
decays. It is clear that both on-shell and off-shell or semi-of-shell vertices,
especially $\pi^{0(*)} \gamma^*\gamma^{(*)}$, play a crucial role in many other
experiments, from the famous $g-2$ (cf. \cite{g2}) via virtual photons stemming
from $e^+e^-$ (see e.g. the recent paper \cite{rosner})
to astrophysics. Our first aim is the common formulation of these interrelated
processes in the given formalism at the given order (either NLO or NNLO)
motivated by the precision of the appropriate experiments.

%%%%%%%%%%%%%%%%%%%%%%%%%%%%%%%%%%%%%%%%%%%%%%%%%%%%%%%%%%%%%%%%%%%%%%%
\section{$\pi^0 \to \gamma \gamma$}

The $\pi^0$ meson has a prominent position among all
hadron particles as being the lightest state of them.  Its primary decay mode is
thus $\pi^0 \to \gamma\gamma$ which is connected with the famous
Adler-Bell-Jackiw triangle anomaly \cite{ABJ}.
It saturates the decay width with almost 99\% and plays an
important role in the further decay modes (see the following sections). The
history of $\pi^0\to\gamma\gamma$ is going back to Steinberger's calculation
\cite{steinberger}, for a review see e.g.
\cite{georgi}, and also \cite{horejsi}. The prediction estimated from the
chiral anomaly using current algebra agrees surprisingly very well with
experiments. A first attempt to explain the small existing deviation from the
measurement was made by Y.~Kitazawa \cite{kitazawa}. At that time a new
experimental prediction from
CERN-NA030 \cite{exp:cern} suggested a smaller value for the partial width
$7.25\pm 0.23$ eV (statistical and systematic errors combined in quadrature).
Older experiments (Tomsk, Desy and Cornell \cite{exp:prim}), seemed not to be so
precise ($7.23\pm0.55$,\ $11.7\pm1.2$,\ $7.92\pm0.42$ eV, respectively); they
relied on the so-called Primakoff effect \cite{primakoff} that is based on
measuring the cross section for the photoproduction of the meson in the Coulomb
field. The more precise number from the direct measurement at CERN motivated
Y.~Kitazawa to explain the $8\sim 9 \%$ discrepancy by including QED correction
and the $\eta/\eta'$ contribution. These corrections were not, however, large
enough to explain the discrepancy which was attributed by the author to a
possible $\pi(1300)$ contribution. Furthermore, it was found out in this work
that the contribution from multi-pion states must be small. This was verified
explicitly also within ChPT with the remarkable observation \cite{donbij} that
at one-loop order there are no chiral logarithms (either from pions or kaons).
The $\pi$-$\eta$-$\eta'$ mixing and electromagnetic correction were reconsidered
relatively recently in \cite{reco}.

The spread in the data basis of the PDG, summarized in the previous paragraph,
shows, however, that the quoted errors seem to be underestimated
\cite{bernstein}. The present situation fortunately looks more optimistic as the
world average accuracy of 8~\% is planned to be improved to the level of one or
two percents in ongoing experiment PrimEx at JLab. The official Run-I result
quoted in \cite{primex} is $\Gamma(\pi^0\to\gamma\gamma) = 7.82\,\text{eV}
\pm2.8\%$. This
was the
main motivation for a new study of $\pi^0\to\gamma\gamma$ in \cite{KM}. The
correction to the chiral anomaly due to the finite mass of light quarks was
reconsidered using strict two-flavour ChPT at NNLO. We will summarize here this
remarkably simple result (note that it involves a two-loop calculation and that
it represents formally a full $O(p^8)$ result). Defining a reduced $T$ amplitude
\begin{equation}
A = e^2 \varepsilon_{\mu\nu\alpha\beta} \epsilon_1^{*\mu} \epsilon_2^{*\nu}
k_1^\mu k_2^\nu\ T \,,
\end{equation}
we have for the partial decay width
\begin{equation}
\Gamma_{\gamma\gamma} = \frac{\pi}{4} \alpha^2 m_{\pi^0}^3 |T|^2\,.
\end{equation}
Up to and including next-to-next-to-leading order corrections
\begin{align}
F_\pi T_{NNLO} &=  {1\over 4\pi^2}
+{16\over3}{m_\pi^2}\left(-4\cwr{3}-4\cwr{7}+\cwr{11}\right)
+{64\over9}{B(m_d-m_u)}(5\cwr{3} + \cwr{7}+2\cwr{8})
\notag\\%%%%%%%%%%%%%%%
&+ {M^4\over 16\pi^2 F^4}\,L_\pi
\left[ {3\over256\pi^4} +
{32F^2\over3}\left(2\cwr{2}+4\cwr{3}+2\cwr{6}+4\cwr{7}-\cwr{11}\right)
\right]
\notag\\%%%%%%%%%%%%%%%
&+{32 M^2 B(m_d-m_u)\over 48\pi^2 F^4}\,L_\pi\,
\left[-6\cwr{2}-11\cwr{3}+6\cwr{4}-12\cwr{5}-\cwr{7}-2\cwr{8}\right]
\notag\\%%%%%%%%%%%%%%%
&-{M^4\over24\pi^2 F^4}\,\left( {1\over16\pi^2}L_\pi \right)^2
+{M^4\over F^4} \lambda_+   +{M^2 B(m_d-m_u) \over F^4} \lambda_-
+{B^2(m_d-m_u)^2\over F^4} \lambda_{--}\ ,
\end{align}
where the chiral logarithm is denoted by $L_\pi=\log{m_\pi^2\over\mu^2}$
and $\lambda_+$, $\lambda_-$, $\lambda_{--}$
can be expressed as follows in terms of renormalized
chiral coupling constants ($d^{Wr}$ refer to combinations of couplings from
the NNLO Lagrangian, i.e. of order $p^8$ in the anomalous sector),
\begin{eqnarray}
&& \lambda_+ = {1\over\pi^2}\left[
-{2\over 3} d_+^{Wr}(\mu) -8c_6^r  -{1\over4}(l_4^r)^2
+{1\over 512\pi^4} \left( -{983\over288} - {4\over3}\zeta(3)
+3\sqrt3\, {\rm Cl}_2(\pi/3)
\right)\right]
\nonumber\\
&& \phantom{\lambda_+ = } +{16\over3} F^2
\left[\,8l_3^r(\cwr{3}+\cwr{7})+l_4^r(-4\cwr{3}-4\cwr{7}+\cwr{11})\right]
\nonumber\\
&& \lambda_-= {64\over9}\left[
d_-^{Wr}(\mu) +F^2 l_4^r\,(5\cwr{3}+\cwr{7}+2\cwr{8}) \right]
\nonumber\\
&& \lambda_{--}= d_{--}^{Wr}(\mu)-128F^2 l_7 (\cwr{3}+\cwr{7})\ ,
\end{eqnarray}
with Riemann zeta and Clausen function: $\zeta(3) = 1.202..$ and ${\rm
Cl}_2(\pi/3) = 1.014..$, respectively.

\begin{figure}[hbt]
 \begin{center}
 \includegraphics{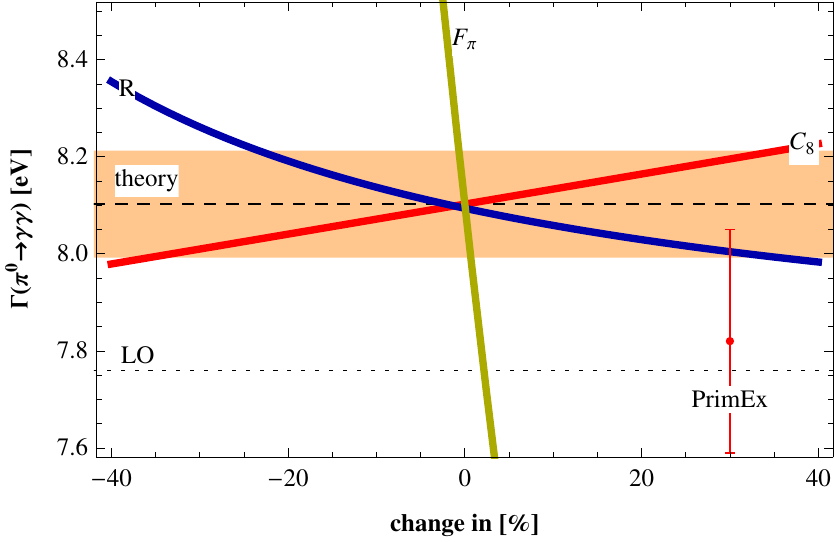}
 \end{center}
 \caption{The dependence of $\pi^0\to\gamma\gamma$ decay width on various
parameters, namely $SU(3)$ LEC $C_8^W$, $R$ and $F_\pi$ (see main text).}
 \label{fig:imp}
\end{figure}

All effects that were carefully studied e.g. in \cite{kitazawa} and \cite{reco}
are now hidden in the LECs and chiral logarithms $L_\pi$ (with the exception of
QED corrections that must be added by hand to the latter formula, see
\cite{KM}). For a detailed phenomenological study of the previous formula
see~\cite{KM}, our best estimate has led to
\begin{equation}
\Gamma_{\gamma\gamma}= (8.09 \pm 0.11)\ {\rm eV }.
\end{equation}
The importance of various input parameters is depicted in Fig.~\ref{fig:imp}.
One can see immediately the
importance of $F_\pi$ in $\pi^0\to\gamma\gamma$ decay. The $F_\pi$, on the other
hand, is determined from the weak decay of $\pi^+$ based on the standard $V-A$
interaction. The new proposed variant of this interaction beyond SM assumes
contributions
of right-handed current which would lead to a change of $F_\pi$ \cite{bernard}.
Determination of this constant directly from $\pi^0$ lifetime can provide
constraints on such contributions. This again put big effort to minimize the
uncertainty stemming from other parameters as for example visualized in
Fig.~\ref{fig:imp}. One of them, a quark mass ratio $R$ will be subject of
Sec.~\ref{sec:eta}, the second one $C_8^W$ reflects the intrinsic connection
with
$\eta\to\gamma\gamma$ decay as calculated in three-flavour ChPT. However,
better understanding of $\eta\to\gamma\gamma$ again in full two-loop
calculation is important and probably necessary as motivated and explained
in~\cite{bk}.

%%%%%%%%%%%%%%%%%%%%%%%%%%%%%%%%%%%%%%%%%%%%%%%%%%%%%%%%%%%%%%%%%%%%%%%
\section{$\pi^0 \to e^+e^- \gamma$}

If $\pi^0\to\gamma\gamma$ represented $99\%$ of all decay modes, $\pi^0\to
e^+e^-\gamma$ represents again more than $99\%$ of the rest modes. It is thus
the second most important decay mode with a branching ratio $\sim 1.174 \pm
0.035\%$ and nowadays called Dalitz decay, after R.H.Dalitz who first
realized its
connection with two-photon production \cite{dalitz}. Knowing the branching ratio
one can also use, at least in principle,
the Dalitz decay to extrapolate the total decay width, which can serve as an
independent possibility how to measure the life-time of $\pi^0$ (however,
handicapped by the larger dependence on error of this ratio). This year's change
of the official PDG number is due to ALEPH archived data on
hadronic $Z$ boson decay which has enabled to reconstruct 12,490 Dalitz decays
with
a result
\begin{equation}
\text{ALEPH:}\qquad\Gamma_{\pi^0\to e^+e^-\gamma}/\Gamma_{\gamma\gamma}  =
(1.140 \pm 0.041)\%,
\end{equation}
that led to the update
\begin{equation}
\text{PDG 10:}\qquad\Gamma_{\pi^0\to e^+e^-\gamma}/\Gamma_{\gamma\gamma}  =
(1.188\pm 0.035)\%.
\end{equation}
To conclude let us also summarize a theoretical prediction:
\begin{equation}\label{eqthd}
\text{theory:}\qquad\Gamma_{\pi^0\to e^+e^-\gamma}/\Gamma_{\gamma\gamma}  =
(1.1851+0.0104+0.0018)\% = (1.1973\pm 0.0055)\%\,,
\end{equation}
where the first number stands for the leading order, predicted already by Dalitz
\cite{dalitz}, the second represents the radiative corrections (numerically
first
done in \cite{joseph}) and the last number \cite{KKN} stands for the prediction
of
QCD corrections and two-photon exchange contribution (which was neglected
previously in QED corrections). The final error was estimated as a half of all
QED corrections. 

The Dalitz decay is very often used experimentally as the normalization mode not
only for rare pion modes (see also below) but also for kaon decay modes, and
thus its precise value has impact on these measurements. Its uncertainty has
in fact a direct effect on external systematic error and different central
values can produce substantial shifts in the final predictions.

The motivation for studying the decay width of $\pi^0 \to e^+e^-\gamma$ is thus
two-fold. The precise and well-understood theoretical prediction with
model-dependent QCD contribution suppressed by phase-space integration can
serve as a calibration of the experiment. On the other hand, if theory is well
under control and still some measurement would signalize some discrepancy the
theory is missing something new and important.

The total decay rate is, however, not the whole story. It turns out that the
corrections to the differential decay which were taken as negligible are indeed
important. The reason is that there is a part of the phase space where,
roughly-speaking, the correction to the differential decay width is positive and
a part where it is negative; and only summing these parts together gives us the
small number in~(\ref{eqthd}). It is clear now, that in physically relevant
applications, when we have to cut some parts of the phase space, these
corrections can become important. Let us discuss this a little bit more in
detail (for details see \cite{KKN}). First it is useful
to define two kinematic variables that represent the normalized di-lepton
invariant mass and the difference of energy for positron and electron
(normalized to
the photon energy in the pion rest frame):
\begin{eqnarray*}
x &=&\frac{(p_+ +p_-)^2}{M_{\pi ^0}^2},\qquad \nu ^2\leq x\leq 1,
\qquad \nu^2\,=\frac{4m^2}{M_{\pi ^0}^2}, \\
y &=&\frac{2P\cdot (p_+ -p_-)}{M_{\pi ^0}^2(1-x)},\,\,\,\,-\sigma
_e(M_{\pi ^0}^2x)\leq y\leq \sigma _e(M_{\pi ^0}^2x),\quad \sigma
_e(s)=\sqrt{1-\frac{4m^2}s}
\end{eqnarray*}
($m$ denotes the electron mass, $p_-^2 = p_+^2 = m^2$ and $P$ stands for the
pion momentum). The next-to-leading corrections to the differential
decay rates can be described as
\begin{equation}
\begin{split}
\frac{\mathrm{d}\Gamma }{\mathrm{d}x\mathrm{d}y} &=\delta(x,y)\,
\frac{
\mathrm{d}\Gamma ^{LO}}{\mathrm{d}x\mathrm{d}y}, \label{ddef} \\
\frac{\mathrm{d}\Gamma }{\mathrm{d}x} &=\delta(x)\,
\frac{\mathrm{d} \Gamma ^{LO}}{\mathrm{d}x}\,,
\end{split}
\end{equation}
where the corresponding LO partial decay rates have a relatively simple form
\begin{align}
\frac{\mathrm{d}\Gamma ^{LO}}{\mathrm{d}x\mathrm{d}y}
&=\frac{\alpha^3}{(4\pi )^4}\frac{M_{\pi^0}}{F_\pi ^2}\,
\frac{(1-x)^3}{x^2}\,[M_{\pi^0}^2 x(1+y^2)+4m^2],\notag \\
\frac{\mathrm{d}\Gamma ^{LO}}{\mathrm{d}x} &=
\frac{\alpha^3}{(4\pi)^4} \frac{8}{3} \frac{M_{\pi^0}}{F_\pi^2}\,
\frac{(1-x)^3}{x^2}\,\sigma_e(x M_{\pi^0}^2) \,(x M_{\pi^0}^2  + 2
m^2) \label{GammaLeading}
\end{align}
(in fact, integrating the last equation one can verify the Dalitz result, i.e.
the first number in (\ref{eqthd})). With these quantities in hand we can
extract information on the QCD part of the form factor ${\cal
F}_{\pi^0\gamma\gamma^{*}}(q^2)$, which is related  to the doubly off-shell
$\pi\gamma\gamma$ transition form factor
defined as
\begin{equation}
\int d^4x\,e^{il\cdot x}\langle 0|T(j^\mu (x)j^\nu (0)|\pi ^0(P)\rangle
%_{c,1\gamma IR}
=-\mathrm{i}\varepsilon ^{\mu \nu \alpha \beta }l_\alpha P_\beta
\,\mathcal{F}_{\pi^0 \gamma^{*} \gamma^{*}}(l^2,(P-l)^2),
\end{equation}
by
\[
\mathcal{F}_{\pi^0 \gamma \gamma^{*}}(q^2) =
\mathcal{F}_{\pi^0 \gamma^{*} \gamma^{*}}(0,q^2).
\]
The Dalitz decay can provides us with the information on the transition form
factor in the time-like region. This is usually specified by its slope parameter
$a_\pi$
\begin{equation}
{\cal F}_{\pi^0\gamma\gamma^{*}}(q^2)\,=\,
{\cal F}_{\pi^0\gamma\gamma^{*}}(0)\,
\big[
1\,+\,a_\pi\,\frac{q^2}{M_{\pi^0}^2}\,+\cdots
\big]\, ,
\end{equation}
and from the experiment can be obtained by subtracting the QED corrections via
\begin{equation}
\frac{\mathrm{d} \Gamma ^{exp}}{\mathrm{d}x}\,-\,
\delta_{QED}(x)\,\frac{\mathrm{d} \Gamma ^{LO}}{\mathrm{d}x}
\,=\,
\frac{\mathrm{d} \Gamma ^{LO}}{\mathrm{d}x}\,
[1 + 2x\,a_\pi],
\label{defa_pi}
\end{equation}

The direct measurements in the time-like region is endowed with large errors
\begin{eqnarray*}
\text{Saclay \cite{exp1}:}\qquad a_\pi &=& -0.11 \pm 0.03 \pm 0.08
\\
\text{TRIUMF \cite{exp2}:}\qquad a_\pi &=& +0.026 \pm 0.024 \pm 0.0048
\\
\text{PSI \cite{exp3}:}\qquad a_\pi &=& +0.025 \pm 0.014 \pm 0.026
,
\end{eqnarray*}
whereas the values extracted from the extrapolation of data at higher energies
in the space-like region, $Q^2=-q^2> 0.5$ GeV$^2$ are more precise:
\begin{eqnarray*}
\text{CELLO \cite{cello}:}\qquad a_\pi &=& +0.0326 \pm 0.0026 \pm 0.0026
\\
\text{CLEO \cite{cleo}:}\qquad a_\pi &=& +0.0303 \pm 0.0008 \pm 0.0009 \pm
0.0012
\end{eqnarray*}
The theoretical calculation \cite{KKN} (here without isospin correction
for simplicity) is given by
\begin{equation}
a_\pi= -\frac{32\pi^2 M_{\pi^0}^2 }{3}c^{Wr}_{13}(M_V^2) -
\frac{M_{\pi^0}^2}{96 \pi^2 F_\pi^2} \Bigl( 1+ 2\ln \frac{M_{\pi^0}^2}{M_V^2}
\Bigr) - \frac{1}{360}\frac{\alpha}{\pi}
\end{equation}
We can see that the EM corrections are very small\footnote{Let us note that
in previous experimental analyses the two-photon contribution were
neglected and their omission would lead to approximately 0.005 correction into
the right direction towards the independent CELLO or CLEO result.} which
signalized its tight connection with a low energy constant $c_{13}^{W}$ (or via
$SU(2)-SU(3)$ relations \cite{KM} to $SU(3)$ LEC $C_{22}^{W}$. The LMD
prediction for
$c_{13}^{W}$ leads to
\begin{equation}
 \text{theory \cite{KKN}:}\qquad a_\pi = +0.029 \pm 0.005
\end{equation}
Let us also note that the off-shell form factor ${\cal F}_{\pi\gamma^*\gamma^*}$
plays an important role in the $\pi^0$-exchange contribution to a hadronic
light-by-light scattering contribution in $g-2$ and thus more
experimental information on this factor can help us to understand the
consistency in different approaches.

%%%%%%%%%%%%%%%%%%%%%%%%%%%%%%%%%%%%%%%%%%%%%%%%%%%%%%%%%%%%%%%%%%%%%%%
\section{$\pi^0 \to e^+ e^- e^+ e^-$}

Having the well established decay of $\pi^0$ into two photons it is clear that
the pion cannot be a
$J=1$ state (so-called Young-Landau theorem \cite{yang}). However, using
directly $\pi^0 \to \gamma\gamma$ to verify whether it is a
(pseudo)scalar is experimentally impossible. It was thus suggested in
\cite{krollwada} to use
the double-internal conversion, the so-called double-Dalitz decay. The
experiment was performed at Nevis Lab \cite{samios} in a bubble chamber with the
following result for the branching ratio:
\begin{equation}
\frac{\Gamma_{e^+e^-e^+e^-}^\text{PDG'08}}{\Gamma_{tot}} = (3.14\pm 0.30)\times
10^{-5}
\end{equation}
which was used as only relevant measurement for almost half century.
The experiment also confirmed the negative parity of $\pi^0$ known from
the previous indirect
studies via the cross-section of $\pi^-$ capture on deuterons. However, the
significance of this direct measurement was only 3.6 $\sigma$. Recently the
long standing experimental gap was filled with a new measurement in the
KTeV-E799 experiment at Fermilab \cite{abouzaid2} giving a branching ratio
(including the radiative final states above a certain cut as tacitly assumed for
all Dalitz modes)
\begin{equation}
\frac{\Gamma_{e^+e^-e^+e^-}^\text{KTeV}}{\Gamma_{tot}} = (3.46\pm 0.19)\times
10^{-5}\,,
\end{equation}
which is in good agreement with the previous experiment. In addition to the
precisely verified parity of $\pi^0$ (which represents its best direct
determination) this experiment sets the first limits on the parity and CPT
violation for this decay. More precisely, having a $\pi^0\gamma^*\gamma^*$
vertex  $C_{\mu\nu\rho\sigma} F^{\mu\nu}F^{\rho\sigma} \pi^0$ we can study,
using the following decomposition (for details see \cite{barker})
$$
C_{\mu\nu\rho\sigma} = \cos\zeta \varepsilon_{\mu\nu\rho\sigma} + \sin\zeta {\rm
e}^{i\delta} (g_{\mu\rho}g_{\nu\sigma} - g_{\mu\sigma}g_{\nu\rho})\,,
$$
the parameters $\zeta$ and $\delta$ which represent parity mixing and CPT
violation parameters. For details see \cite{abouzaid2}; for example their limit
on the mixing assuming CPT conservation is $\zeta<1.9^\circ$.

A detailed analysis of the radiative corrections in \cite{barker} showed that
they seem to be very important in extracting physically relevant quantities.
This motivates us to reopen this subject \cite{KKN2} in the same manner as was
done in \cite{KKN}. The simply looking task of attaching another Dalitz pair on
the virtual photon line is complicated (in the defined power-counting) by the
necessity to include a pentagonal diagram \cite{barker}. This strengthens the
need of a correct description of the off-shell $\pi^0\gamma^*\gamma^*$ vertex,
which can be, on the other hand, directly studied in the next mode.

%%%%%%%%%%%%%%%%%%%%%%%%%%%%%%%%%%%%%%%%%%%%%%%%%%%%%%%%%%%%%%%%%%%%%%%
\section{$\pi^0 \to e^+ e^-$}
In the previous decay modes the fully off-shell $\pi^0 \gamma^* \gamma^*$
vertex was suppressed by the dominant semi-on-shell contributions. As this is
not true anymore for $\pi^0\to e^+ e^-$ it naturally represents the simplest
and cleanest candidate for studying not-well understood effects of QCD, i.e.
$\mathcal{F}_{\pi^0 \gamma^* \gamma^*}(k^2\neq 0,l^2 \neq 0)$, even though the
process itself is suppressed by approximate helicity conservation and two extra
powers of $\alpha$.
This is supported by the existing experiment at Fermilab (KTeV E799-II)
\cite{abouzaid1}. Comparing with the previous measurements their result has
increased significantly the precision and provide thus the most important
contribution to the present PDG's average
$$
\frac{\Gamma_{e^+e^-}^\text{PDG}}{\Gamma_{tot}} = (6.46 \pm 0.33)\times
10^{-8}\,.
$$
This process was first calculated in \cite{drell} and proceeds, as mentioned,
via
two intermediate photons and at LO is thus represented by an one-loop
(triangle) diagram. One can get (for details see e.g. \cite{kppr})
\begin{equation}
 \frac{\Gamma_{e^+e^-}}{\Gamma_{\gamma\gamma}} = 2 \Bigl(\frac{\alpha m}{\pi
M_{\pi^0}}\Bigr)^2 \sigma(M_{\pi^0}^2) |{\cal A}(M_{\pi^0}^2)|^2\,,
\end{equation}
with the amplitude given by
\begin{equation}
 {\cal A}(s) = \chi(\mu) - \frac52 + \frac32 \ln \frac{m^2}{\mu^2} + C(s)\,,
\end{equation}
where $C$ represents the scalar one-loop triangle. The imaginary part can be
calculated in a model independent way by
cutting the photon lines and knowing the on-shell form factor
$\mathcal{F}_{\pi^0 \gamma^* \gamma^*}(0,0)$ (in fact as we have a normalization
to
two-photon decays it has to be equal 1) with the result
\begin{equation}
 \text{Im}\,{\cal A} = \frac{\pi}{2\sigma}\ln\frac{1-\sigma}{1+\sigma}\,,
\end{equation}
implying thus a unitarity bound on the ratio
$\Gamma_{e^+e^-}/\Gamma_{\gamma\gamma} \gtrsim 5\times 10^{-8}$. The real part
can
be now calculated via a dispersive integral leaving us with one unknown
subtraction, or equivalently by $\chi(\mu)$. The techniques of
large $N_C$ together with the LMD approximation ($V$ represents
the $\rho$ meson) leads to \cite{kppr}
\begin{equation}
 \chi(\mu = M_V)  = \frac{11}{4} - 4\pi^2 \frac{F_0^2}{M_V^2} = 2.2\pm 0.9
\quad\Rightarrow\quad
\frac{\Gamma_{e^+e^-}^\text{th.no-rad}}{\Gamma_{\gamma\gamma}} = (6.2 \pm
0.3)\times
10^{-8}\,.
\end{equation}
This should be compared with the experiment after removing the effects of final
state radiation:
\begin{equation}
  \frac{\Gamma_{e^+e^-}^\text{KTeV.no-rad}}{\Gamma_{\gamma\gamma}} =
(7.57\pm0.39)\times 10^{-8}\ ,
\end{equation}
which is $3.5\sigma$ off from the mentioned theoretical prediction.

At present the theoretical activities concerning this process focus on two
main directions: {\it i\/}) understanding the discrepancy within the SM -- i.e.
calculating radiative corrections or employing a different model for the QCD
part
\cite{dorokhov}. {\it ii\/}) It naturally represents an ideal candidate for
testing new models
beyond SM, it can set valuable limit on light dark matter scenarios,
supersymmetric extensions (axion), etc...

Before concluding let us also mention the weak contribution. This process has
in fact a direct tree level contribution mediated via $\pi^0 \to Z^* \to
e^+e^-$. The amplitude would be proportional to the Fermi constant:
\begin{equation}
 {\cal A}^\text{weak}_{\pi^0\to e^+e^-}   \sim G_F F_\pi m \bar u \gamma_5
v\,,
\end{equation}
which makes it three orders of magnitude smaller than the dominant EM
contribution.
Beyond-SM scenarios would use a similar relation (with different coupling
$G_F$) for introducing the effect of a light vector particle (e.g. U-boson).
%%%%%%%%%%%%%%%%%%%%%%%%%%%%%%%%%%%%%%%%%%%%%%%%%%%%%%%%%%%%%%%%%%%%%%%
\section{$\pi^0 \to \nu \bar \nu $, invisible, extra-light particles}
\label{secnunu}

As studied for example by \cite{LamNg} there is a tight connection for
$\pi^0\to\nu\bar\nu$ with cosmology so that the strong limit on
this decay obtained are much higher than those in the laboratory:
$2.7\times 10^{-7}$ (E949 based on $K^+\to \pi^+\pi^0$ \cite{e949}).
This decay mode represents not only two neutrino decay modes but all possible
combination and weakly coupled exotics and also even more generally
$\pi^0 \to \text{invisible}$.

Within the SM (extended by massive neutrinos) one can use the calculation of
the weak sector for electron from previous section with a prediction
\begin{equation}
 {\cal A}^\text{weak}_{\pi^0\to \nu^+\nu^-} = \sqrt2 G_F F_\pi m_\nu \bar u
\gamma_5
v\,,
\end{equation}
with one subtlety -- we don't know the mass of the neutrino and its nature (for
the Majorana type the amplitude is twice bigger). The relative branching ratio
is then (in the Dirac case)
\begin{equation}
 \frac{\Gamma_{\nu\bar\nu}}{\Gamma_{\gamma\gamma}} = \Bigl( \frac{4\pi F_\pi^2
G_F}{\alpha} \frac{m_\nu}{M} \Bigr)^2 \sqrt{1-\frac{4m_\nu^2}{M^2}}
\end{equation}
With a direct limit on the absolute tau neutrino mass $m_\nu^\tau < 18.2$ MeV we
get a reasonably high limit on branching ratio $< 5\times 10^{-10}$ (and twice
bigger for Majorana case). Note the maximum for the ratio for $m_\nu =
M/\sqrt6$, which is, however, ruled out.

The helicity suppression for $\pi^0 \to \nu \bar \nu$
can be avoided in the decay mode $\pi^0\to \nu \bar\nu\gamma$. This decay mode
is also interesting as it depends on the actual number of light neutrinos.

A general possibility to set the constraints on an extra-light long-lived
(/non-interacting/weakly interacting) neutral vector particle $X$ via decays of
known particles opens naturally this question also for the exotic
$\pi^0$ decay.

%%%%%%%%%%%%%%%%%%%%%%%%%%%%%%%%%%%%%%%%%%%%%%%%%%%%%%%%%%%%%%%%%%%%%%%
\section{$\eta\to 3\pi$}
\label{sec:eta}

We have selected this representative of $\eta$ decay modes as the most
important example of the hadronic decays for the studied particles. A brief look
into the history of this mode (\cite{gl} for NLO and \cite{bg} for NNLO)
shows even after two-loop calculation a discrepancy between theory and
experiments. This could be attributed either to bad convergence of the ChPT
series
and necessity to somehow (at least partially) re-sum higher orders or to some
problems in the calculation at NNLO (e.g. to a wrong estimate of LECs). The new
ongoing efforts represent in some sense combinations of these two possibilities
(\cite{zkkn}, \cite{cle}, \cite{skd}).

Apart from the obvious motivation to better understand how to perform NNLO
calculation in ChPT in the three flavour case, which seems to be a problematic
subject we have in hands a process which vanishes in the isospin limit. It means
that a precise measurement together with a good understanding of the theory
should provide us with parameters which can quantify the isospin breaking, for
example:
$$
R = \frac{m_s- \hat m}{m_d-m_u}\qquad Q^2 = \tfrac12 \Bigl(1+ \frac{m_s}{\hat
m}\Bigr) R\,.
$$

The absolute value of the partial decay width for $\eta\to3\pi$ is
experimentally obtained via normalization to $\eta\to\gamma\gamma$. A change in
one decay width has thus influence in other (a change by 1\% in
$\Gamma(\eta\to\gamma\gamma)$ input shifts R by $\approx$ 0.2).

%%%%%%%%%%%%%%%%%%%%%%%%%%%%%%%%%%%%%%%%%%%%%%%%%%%%%%%%%%%%%%%%%%%%%%%
\section{Summary}
The new experimental activities in the low energy physics that concern directly
$\pi^0$ or $\eta$ decay modes call for a more detailed theoretical study in this
area. We have discussed some allowed decay modes that
represent important tools for studying basic
phenomena of the underlying theory: QCD. Namely, $\pi^0\to\gamma\gamma$ and
$\eta\to\gamma\gamma$
played an important role in understanding a symmetry pattern of the theory as
they are directly connected with the so-called $U(1)_A$ anomaly. QCD enlarged
by photons possesses, however, two such anomalies. The first one, internal,
connected with QCD only, proportional to gluonic term $G_{\mu\nu}\tilde
G^{\mu\nu}$, dubbed $U(1)$-problem and the resulting strong $CP$ problem is
still
an open issue. As a remnant of the anomaly, the $\eta'$
plays a more important role than naively expected and has to be included in a
theoretical consideration. The second anomaly, external, in our case connected
with
electromagnetic interaction (or $F_{\mu\nu}\tilde F^{\mu\nu}$) explains why
$\pi^0\to\gamma\gamma$ can decay so quickly even it should be suppressed due
to Sutherland's theorem. Furthermore, $\eta\to\gamma\gamma$ and $\eta \to 3\pi$
represent 95\% of
all $\eta$-decay modes and are thus perfectly suited to study directly
properties of the $\eta$.
Simultaneous treatment of two-photon $\pi^0$ and $\eta$ decays, apart from
testing or fixing our understanding of $\eta'$, can provide valuable information
on the decay constants $F_\pi$ and $F_\eta$ or quark mass ratio.

A common treatment
is thus useful and important in order to understand all phenomena. In this work
we have mainly focused on chiral and QED corrections in order to prepare the
ground for
the discussion of non-perturbative effects or eventual new physics.



%% References


%\bibliographystyle{elsarticle-num}






% \end{thebibliography}

\end{document}